\documentclass{article}
\usepackage{latexsym}
\usepackage{graphics}
\newtheorem{proposition}{Proposition}
\newtheorem{theorem}{Theorem}
\newtheorem{lemma}{Lemma}
\newtheorem{corollary}{Corollary}

\newcommand{\tl}{\frac{\theta}{\lambda}}
\newcommand{\zn}{\frac{\zeta}{\nu}}
\begin{document}
\author{\sc Mihalis Dafermos\footnote{supported in part by
the NSF grant DMS-0302748} \\ Department of Mathematics, MIT \\
dafermos@math.mit.edu}
\title{Black hole formation from a complete regular past}
\maketitle
\begin{abstract}
An open problem in general relativity has been to construct
an asymptotically flat solution to a reasonable
Einstein-matter system containing a black hole in the future
and yet past-causally geodesically complete, in particular, containing no white 
holes. We give such an example in this paper--in fact a family of
such examples, stable in a suitable sense--for 
the case of a self-gravitating scalar field.
\end{abstract}
\section{Introduction}
The problem of gravitational collapse is typically formulated as
the study of the future maximal evolution of complete asymptotically
flat Cauchy data for an appropriate Einstein-matter system. 
The question of identifying physically admissible initial data,
however, is best characterized by properties of their \emph{past} 
evolution. In particular, it seems reasonable to restrict
to Cauchy data whose past evolution is regular.
Unfortunately, however, with the exception of certain classical results
for dust~\cite{os:?},
current theorems on the evolution of asymptotically flat Cauchy data 
ensuring a regular past~\cite{ck:book, chr:sgsf,rr:ge} 
also ensure a regular future; this is due to
 the fact that these theorems depend
on smallness in function-space norms that
do not distinguish past from future.\footnote{One can compare
with the cosmological case, where choice of gauge based on
the expansion of the universe can be
used to prove future completeness theorems
in cases where the past is known to be singular. See for 
instance~\cite{arr:ohs,
gr:fgc}.}
As a consequence, even a single example of a solution to a reasonable 
Einstein-matter system
with a regular asymptotically flat past and a singular future has thus
 far been lacking.

In the present paper, we shall prove
\begin{theorem}
There exist past-causally 
geodesically complete solutions of the coupled
Einstein-scalar field equations, 
which to the future terminate in a $C^0$-singularity hidden inside a 
black hole. These solutions are maximal developments of complete
asymptotically flat Cauchy data. 
\end{theorem}
In fact, a large class of such solutions will be proven to exist.
The solutions are spherically symmetric and
their conformal diagram\footnote{This is defined to be the image
of a conformal representation of the quotient manifold $Q=M/SO(3)$
in a bounded domain of $2$-dimensional Minkowski space.} is as follows:
\[
\includegraphics{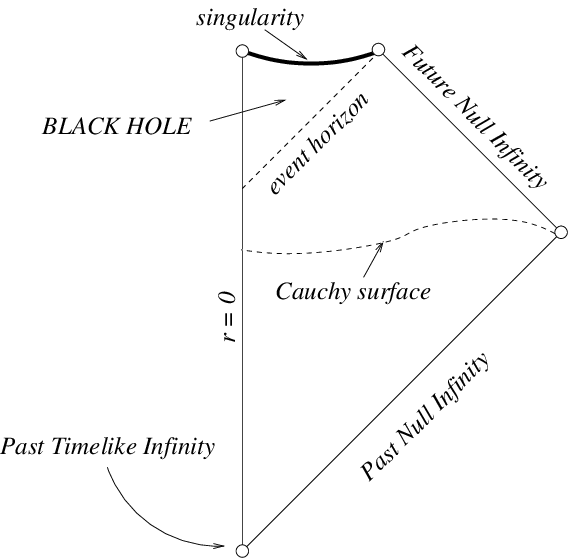}
\]
Moreover,
the above picture will be stable to a certain class of perturbations.

Although the solutions we construct will admit complete asymptotically
flat Cauchy surfaces, they will not be constructed
\emph{a priori} as developments of such data. Rather, the examples
of this paper will be constructed by
pasting together the solutions to three distinct characteristic
initial value problems. In particular,
the issue of formulating a criterion
on Cauchy data ensuring a singular future yet regular
past is sidesteped in this paper. 

Replacing the Cauchy problem with several distinct
characteristic initial value problems allows
one to completely separate the problem of the future from the past.
The future is determined by data prescribed on a future-outgoing cone (see
$A$ in the diagram below),
whereas the past is determined by data on a past-outgoing cone (see $B$).
To complete the spacetime, however, one has to construct (and understand)
the solution ``between'' the two cones (see $\Gamma$):
\[
\includegraphics{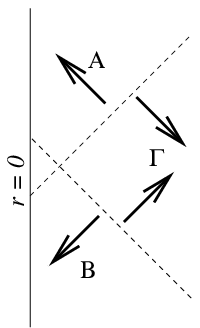}
\]
Since in spherical symmetry, the equations reduce to a quasilinear
system in $2$-dimensions, the problem depicted by
$\Gamma$ is also a well-posed characteristic initial value problem.
Conceptually, this 
is perhaps easiest understood by replacing the two-dimensional metric
on $Q$ by its negative, thus interchanging the notions of space and 
time.\footnote{Estimates in the direction depicted by $\Gamma$ have many
applications in the study of $2$-dimensional hyperbolic problems. See
\cite{md:tp, dr:dr, cs:wm, chr:mt}.}

The data for the initial value
problem $B$ can be taken to be trivial, but alternatively,
from Christodoulou~\cite{chr:sgsf, chr:bv}, one can 
prescribe an arbitrary scalar field sufficiently small in an appropriate
norm, and this
will also guarantee a past complete development. On the other hand,
Christodoulou has given in~\cite{chr:fbh}
a criterion on initial data ensuring that
for the initial value problem $A$, a black hole forms 
in the future. The important feature, from
the point of view of this paper, is that this criterion can be applied
to data which are not trapped initially, and for which the mass aspect
function $\mu$ and the scalar field multiplied by the area radius 
$r\phi$ can be arbitrarily small in the supremum norm.

Theorem 1 thus reduces to completely understanding the evolution
of problem $\Gamma$. It should be noted that the global features of
this problem are quite different from the more standard characteristic initial
value problems $A$ and $B$; in particular, 
there are no \emph{a priori} global energy
bounds. The main analytical result of this paper is
that $\Gamma$ leads to a ``complete'' wedge in an asymptotically flat
spacetime, given sufficiently small initial data. The requisite
smallness, however, is asymmetric in the two initial characteristics;
on the future outgoing characteristic, only pointwise 
smallness of $\mu$ and $r\phi$
are required. In particular, the requirements for the data for $\Gamma$
are compatible with the data necessary in $A$ to produce a black hole. 

For completeness, in the next section,
the Einstein-scalar field equations will be presented under
spherical symmetry. The initial value problem $\Gamma$ is studied in
Section 3. From this, the theorem of this paper is easily obtained
in Section 4 along the considerations outlined above.
\vskip1pc
\noindent{\bf Acknowledgement:} The importance of identifying asymptotically
flat initial data whose future development is singular but
whose past development is regular has been stressed by Sergio Dain and Alan
Rendall. I thank them for several very useful discussions.

\section{The Einstein-scalar field equations under spherical
symmetry}
For a detailed discussion of the role of the
Einstein-scalar field system in the study of
the problem of gravitational collapse, the reader may refer to
\cite{md:si}.
In rationalized units, the equations take the form
\[
R_{\mu\nu}-\frac12Rg_{\mu\nu}=2T_{\mu\nu},
\]
\[
g^{\mu\nu}\phi_{;\mu\nu}=0,
\]
\[
T_{\mu\nu}=\phi_{;\mu}\phi_{;\nu}
-\frac12g_{\mu\nu}\phi^{;\alpha}\phi_{;\alpha},
\]
where $g_{\mu\nu}$ is a Lorentzian metric defined on a $4$-dimensional
manifold $M$ and $\phi$ is a scalar function (the scalar field).

Recall that spherical symmetry is the assumption that
$SO(3)$ acts by isometry on the spacetime and preserves $\phi$.
Under this assumption,
the equations reduce to a second order system for 
functions $(r,g_{ij},\phi)$ 
defined on the space of group orbits $Q$ of
the $SO(3)$ action. Here $r$ is the so-called \emph{area radius} function,
i.e.~$r$ evaluated at a point $p$ of $Q$ retrieves up 
to a constant the square root of
the area of the group orbit corresponding to $p$ in $M$. This group orbit
is necessarily a spacelike sphere. The tensor $g_{ij}$ is a 
$1+1$-dimensional Lorentzian metric on $Q$, which is induced from
the metric on $M$. The functions
$r$ and $g_{ij}$ together determine $g_{\mu\nu}$ as follows:
\[
g_{\mu\nu}dx^\mu dx^\nu=g_{ij}dy^idy^j+r^2\gamma,
\]
where $\gamma$ is the standard metric on the unit $2$-sphere.
Since the scalar field is constant on group orbits, it descends
to a function on $Q$, and will still be denoted by $\phi$.

One can prescribe null coordinates $u$ and $v$ on $Q$ so that its metric
becomes $-\Omega^2dudv$. These coordinates then represent
the characteristic directions of the equations. 
To exploit the method of characteristics, it
is more convenient to write the equations as a first order system.
Moreover, $\Omega$ can be replaced as an unknown in the equations
by the Hawking mass 
\[
m=\frac{r}2(1-|\nabla r|^2).
\]
Introducing
the first order derivatives $\partial_ur=\nu$, $\partial_vr=\lambda$,
$r\partial_u\phi=\zeta$, $r\partial_v\phi=\theta$, and also
the so-called mass aspect function $\mu=\frac{2m}r$, we obtain the
system
\begin{equation}
\label{ruqu}
\partial_u{r}=\nu,
\end{equation}
\begin{equation}
\label{rvqu}
\partial_v{r}=\lambda,
\end{equation}
\begin{equation}
\label{lqu}
\partial_u\lambda=\lambda\left(\frac{2\nu}{1-\mu}\frac{m}{r^2}
\right),
\end{equation}
\begin{equation}
\label{nqu}
\partial_v\nu=\nu\left(\frac{2\lambda}{1-\mu}\frac{m}{r^2}
\right), 
\end{equation}
\begin{equation}
\label{puqu}
\partial_u m=\frac{1}{2}(1-\mu)\left(\zn\right)^2\nu,
\end{equation}
\begin{equation}
\label{pvqu}
\partial_v m=\frac{1}{2}(1-\mu)\left(\tl\right)^2\lambda,
\end{equation}
\begin{equation}
\label{sign1}
\partial_u\theta=-\frac{\zeta\lambda}r,
\end{equation}
\begin{equation}
\label{sign2}
\partial_v\zeta=-\frac{\theta\nu}r.
\end{equation}

From the above we also derive
\begin{equation}
\label{avafora}
\partial_v(\phi\nu+\zeta)=\frac{\phi}{r^2}\frac{2\lambda\nu}{1-\mu}
				m.
\end{equation}
\begin{equation}
\label{avafora2}
\partial_u(\phi\lambda+\theta)=\frac{\phi}{r^2}\frac{2\lambda\nu}{1-\mu}
				m.
\end{equation}
We shall see that the $r^{-2}$ factor on the right hand side above
allows us to show that the quantities $\phi\lambda+\zeta$,
$\phi\lambda+\theta$
have better decay in $r$ than $\phi$, $\theta$, or $\zeta$; this 
fact plays an important role in our argument. 

We also easily obtain from the above the equations
\begin{equation}
\label{fdb1}
\partial_u\frac{\lambda}{1-\mu}=
\frac1{r}\left(\zn\right)^2\nu\frac{\lambda}{1-\mu}
\end{equation}
and
\begin{equation}
\label{fdb2}
\partial_v\frac{\nu}{1-\mu}=
\frac{1}r\left(\tl\right)^2\lambda\frac{\nu}{1-\mu}.
\end{equation}

\section{The characteristic initial value problem $\Gamma$}
In this section, we will study
a characteristic initial value problem posed
in the direction indicated by $\Gamma$ in the Introduction.
This will be the main new analytic element in the
proof of Theorem 1 in
the next section.

For the time being, fix constants $R>0$, $C>0$, $M>0$, and 
$m_0\ge0$.
The initial characteristic segments will be $u=-R$ and $v=R$.
At $(-R,R)$, we prescribe 
\[
r(-R,R)=R, 
\]
\[
m(-R,R)=m_0.
\]
On $u=-R$, we prescribe $\mathcal{C}^2$ 
functions of $v$ $r(-R,v):[R,\infty)\to{\bf R}$, 
$\phi(-R,v):[R,\infty)\to{\bf R}$ so as to satisfy
\begin{equation}
\label{vd1}
\partial_vr=\lambda=1,
\end{equation}
\begin{equation}
\label{vd2}
|r\phi|\le C,
\end{equation}
\begin{equation}
\label{vd3}
m\le M.
\end{equation}
On $v=R$, we prescribe $\mathcal{C}^2$ functions 
$r(u,R):(-\infty,-R]\to{\bf R}$, 
and $\phi(u,R):(-\infty,-R]\to{\bf R}$, so as to satisfy
\begin{equation}
\label{ud1}
\partial_ur=\nu=-1, 
\end{equation}
\begin{equation}
\label{ud2}
|\partial_u(r\phi)|=|\zeta+\nu\phi|\le C|u|^{-2},
\end{equation}
\begin{equation}
\label{ud3}
m\le M.
\end{equation}

We have the following
\begin{proposition}
\label{LWP}
Consider the initial value problem described above.
There exists a unique subset ${\bf D}\subset(-\infty,-R]\times[R,\infty)$, 
open in the subspace topology of the latter set,
and a unique set of sufficiently regular functions $r$, $\phi$, and $m$
defined on ${\bf D}$, such that
\begin{enumerate}
\item
$-R\times[R,\infty)\cup(-\infty,-R]\times R\subset {\bf D}$ and
$r$, $\phi$, $m$ match with the data prescribed.
\item
The functions $r$, $\phi$, $m$ satisfy $(\ref{ruqu})$--$(\ref{sign2})$.
\item
If $(u,v)\in{\bf D}$, then $(\tilde{u},\tilde{v})\in{\bf D}$,
for all $\tilde{u}\ge u$, $\tilde{v}\le v$.
\item
Given any other $\tilde{\bf D}$, and sufficiently regular
$\tilde{r}$, $\tilde{m}$, $\tilde{\phi}$,
satisfying properties 1, 2, and 3, it follows that
$\tilde{\bf D}\subset {\bf D}$,
and $\tilde{r}=r$, $\tilde{m}=m$, and $\tilde{\phi}=\phi$ on 
$\tilde{\bf D}$.
\end{enumerate}
Moreover, if the initial data are $C^\infty$, then 
$r$, $\phi$, $m$ are $C^\infty$ functions on ${\bf D}$.
\end{proposition}

\noindent\emph{Proof.} The proof is by standard techniques and
is omitted. $\Box$

We shall call ${\bf D}$ the maximal development of 
our initial value
problem. Time orienting the negative of the induced metric 
$-g_{ij}$ of our solution in the
obvious way, property 2 above says that ${\bf D}$ is a past set
with respect to $-g$. In what follows, we will 
denote causal relations with respect to the 
negative metric under the
above time orientation by $J^+_{-g}$, $J^{-}_{-g}$,
etc.

\begin{proposition}
For the intial value problem described
above, $\nu<0$ and $\lambda>0$ throughout ${\bf D}$.
\end{proposition}

\noindent\emph{Proof.} 
By continuity, the above inequalities indeed
 hold in a neighborhood of initial data.
Thus, assuming the proposition is false, there exists a
$(u,v)$ such that either $\nu(u,v)=0$ or $\lambda(u,v)=0$, but
for which $\nu<0$, $\lambda<0$ in $J^{-}_{-g}(u,v)$:
\[
\includegraphics{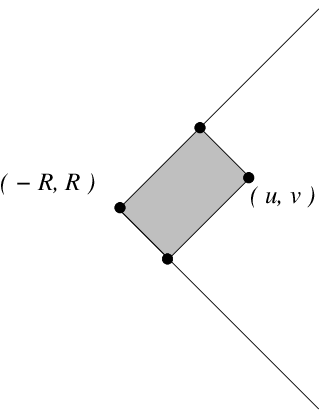}
\]
Let us first assume that $\lambda(u,v)=0$.
Integrating equation $(\ref{fdb1})$ on $[u,-R]\times\tilde{v}$
for each $\tilde{v}\in[R,v]$, we 
obtain that 
\begin{eqnarray*}
\frac\lambda{1-\mu}(u,\tilde{v})	&=&		\frac\lambda{1-\mu}(-R,\tilde{v})
										e^{\int_{-R}^u\frac{1}{r}
										\frac{\zeta^2}
										\nu d\tilde{u}}\\
							&=&		\frac\lambda{1-\mu}(-R,\tilde{v})
										e^{-\int_{u}^{-R}\frac{1}{r}
										\frac{\zeta^2}
										\nu d\tilde{u}}\\
							&\ge&	\frac\lambda{1-\mu}(-R,\tilde{v}),
\end{eqnarray*}
since $\nu\le0$ in $\overline{J^{-}_{-g}(u,v)}$.
It then follows from the above that we have $1-\mu\ge0$ in
$\overline{J^{-}_{-g}(u,v)}$.

Integrating now $(\ref{lqu})$ on $[u,-R]\times v$, we obtain
\begin{eqnarray*}
\lambda(u,v)	&=&		\lambda(-R,v)e^{\int_{-R}^u{\frac{2\nu}{1-\mu}
					\frac{m}{r^2}d\tilde{u}}}\\
		&\ge&	\lambda(-R,v)\\
		&\ge&	1.
\end{eqnarray*}
But this contradicts the assumption $\lambda(u,v)=0$.
In an entirely similar fashion, integrating $(\ref{fdb2})$,
the assumption $\nu(u,v)=0$ leads
to a contradiction. The proposition is thus proven. $\Box$

\begin{proposition}
In ${\bf D}$, 
\begin{equation}
\label{rkaiR}
r\ge R,
\end{equation}
\[
m\ge m_0,
\]
\begin{equation}
\label{nu1}
\nu\le-1,
\end{equation}
\begin{equation}
\label{la1}
\lambda\ge1,
\end{equation}
\begin{equation}
\label{1-mu1}
0<1-\mu\le1.
\end{equation}
\end{proposition}
 
\noindent\emph{Proof.} The proof is immediate from the previous proposition
and the signs in the equations $(\ref{ruqu})$--$(\ref{pvqu})$.
$\Box$

Let $\partial{\bf D}$ denote the boundary of ${\bf D}$ in the topology
of $(-\infty,-R]\times[R,\infty)$. By monotonicity, $r$ and $m$
extend to continuous functions on ${\bf D}\cup\partial{\bf D}$ valued
in the extended real numbers, and causality relations can still be
applied. We have the following extension proposition:

\begin{proposition}
\label{epektasn}
If $p\in{\bf D}\cup\partial{\bf D}$, $J^-_{-g}(p)\subset{\bf D}$,
and $r(p)<\infty$, $m(p)<\infty$, then $p\in{\bf D}$.
\end{proposition}

\noindent\emph{Proof} omitted. $\Box$ 

The achronal (with respect to $-g$) structure of the boundary
$\partial{\bf D}$ immediately yields the following 

\begin{corollary}
If $\partial{\bf D}\ne\emptyset$ then there exists a 
$p\in\partial{\bf D}$ such that either $r(p)=\infty$ or $m(p)=\infty$.
\end{corollary}

We now proceed to the main result of this section:

\begin{proposition}
\label{kurioprop}
Fix $\alpha>1$. Consider an initial value problem of the type described 
before, where\footnote{We have here not attempted to give a scale-invariant
condition.}
\[
R>\max\left\{\left(\frac{4\alpha}{\alpha-1}\right)^{2},
\left(\frac{1+16\alpha^4M}{\alpha}\right)^2,
\frac{8\alpha^2M}{\log \alpha},\frac{6\alpha^2C^2}{\alpha-1},\frac{4\alpha M}
{1-\alpha}\right\}.
\]
Then
\[
{\bf D}=(-\infty,-R]\times[R,\infty),
\]
and the following estimates
hold:
\begin{equation}
\label{E1}
|r\phi|\le \alpha C,
\end{equation}
\begin{equation}
\label{E2}
|\nu\phi+\zeta|\le (1+\alpha^4M)C|u|^{-2},
\end{equation}
\begin{equation}
\label{E3}
-1\ge\nu\ge-\alpha
\end{equation}
\begin{equation}
\label{E4}
1\le\lambda\le\alpha
\end{equation}
\begin{equation}
\label{E5}
1-\mu\ge (2\alpha)^{-1}
\end{equation}
Moreover, if in addition 
\begin{equation}
\label{AA1}
|\lambda\phi+\theta|(R,v)\le\tilde{C}v^{-2},
\end{equation}
then 
\begin{equation}
\label{E6}
|\lambda\phi+\theta|(R,v)\le(\tilde{C}+\alpha^4MC)v^{-2}
\end{equation}
holds as well.
\end{proposition}
\noindent\emph{Proof.} 
Consider the region $\mathcal{R}$ 
defined to be the set of all $p$ such that
the inequalities:
\begin{equation}
\label{BS1}
|r\phi|<2\alpha C
\end{equation}
\begin{equation}
\label{BS2}
|\nu\phi+\zeta|<2\alpha C|u|^{-\frac32}
\end{equation}
\begin{equation}
\label{BS3}
\nu>-2\alpha
\end{equation}
\begin{equation}
\label{BS4}
\lambda<2\alpha
\end{equation}
\begin{equation}
\label{BS5}
1-\mu>(2\alpha)^{-1}
\end{equation}
\begin{equation}
\label{BS6}
m<2\alpha M
\end{equation}
hold for all $q\in J^{-}_{-g}(p)$. With the exception of
$(\ref{BS2})$ and 
$(\ref{BS5})$, the above inequalities hold on the intial data
segments regardless of the value of $R$. Inequality $(\ref{BS2})$ holds since $R>1$, and $(\ref{BS5})$ holds
since
\[
R>\frac{4M}{1-\alpha^{-1}}.
\]
By Proposition \ref{LWP}, $\mathcal{R}$ is non-empty, 
and contains a 
neighborhood of the initial segments; moreover, it is easily seen 
to be open
in the topology of ${\bf D}$.
We shall show first that in $\overline{\mathcal{R}}$, 
$(\ref{BS1})$--$(\ref{BS6})$ hold
with $\alpha$ replacing $2\alpha$. 

Let $(u,v)$ thus be in $\overline{\mathcal{R}}$. It follows by continuity that
$(\ref{BS1})$--$(\ref{BS6})$ hold at $(u,v)\cup J^{-}_{-g}(u,v)$,
where, however, the $<$ sign is replaced by the $\le$ sign.

We estimate $r\phi(u,v)$ as follows. Integrating the equation
\[
\partial_u(r\phi)=\nu\phi+\zeta,\]
we obtain from $(\ref{vd2})$ and $(\ref{BS2})$ 
\begin{eqnarray*}
|r\phi(u,v))|	&\le&	|r\phi(-R,v)|+\int_{u}^{-R}{|\nu\phi+\zeta|d\tilde{u}}\\
	&\le&	C+4C\alpha R^{-\frac12}\\
	& = &	C(1+4R^{-\frac12}\alpha)\\
	& < &	\alpha C
\end{eqnarray*}
since $R>\left[4\alpha(\alpha-1)^{-1}\right]^2$.

On the other hand, integrating $(\ref{avafora})$ and applying
$(\ref{ud2})$, $(\ref{BS1})$, $(\ref{BS3})$, $(\ref{BS5})$, $(\ref{BS6})$, 
we estimate
\begin{eqnarray*}
|\nu\phi+\zeta|(u,v)	&\le&	|\nu\phi+\zeta|(u,R)+
				\int_R^v{\frac{|\phi|}{r^2}
				\frac{2\lambda|\nu|}{1-\mu}m
				d\tilde{v}}\\
		&\le&	C|u|^{-2}+\int_R^{r(u,v)}{\frac{|\phi|}{r^2}
				\frac{2|\nu|}{1-\mu}m
				dr}\\
		&\le&	C|u|^{-2}+32C\alpha^4M
				\int_R^{r(u,v)}{\frac{dr}{r^3}}\\
		&\le&	C|u|^{-2}+16C\alpha^4Mr(u,R)^{-2}\\
		& = &	[C+16C\alpha^4M]|u|^{-2}\\
		&\le&	CR^{-\frac12}[1+16\alpha^4M]
				|u|^{-\frac32}\\
		& < &	\alpha C|u|^{-\frac32},
\end{eqnarray*}
where for the last inequality we use that
$R>\alpha^{-2}(1+16\alpha^4M)^2$.

To estimate $\nu(u,v)$, we integrate $(\ref{nqu})$, applying $(\ref{ud1})$,
$(\ref{BS3})$, $(\ref{BS5})$, $(\ref{BS6})$, to obtain
\begin{eqnarray*}
\nu(u,v)	&\ge&	\nu(u,R)e^{\int_R^v{\frac{2}{r^2}
			\frac{\lambda|\nu|}{1-\mu}md\tilde{v}}}\\
	&\ge&	-e^{8\alpha^2M\int_R^{r(u,v)}{\frac{dr}{r^2}}}\\
	&\ge&	-e^{8\alpha^2MR^{-1}}\\
	& > &	-\alpha
\end{eqnarray*}
since $R> 8\alpha^2 M(\log\alpha)^{-1}$.
Similarly, from $(\ref{lqu})$ we estimate $\lambda(u,v)$,
applying $(\ref{vd1})$, $(\ref{BS4})$, $(\ref{BS5})$, $(\ref{BS6})$: 
\begin{eqnarray*}
\lambda(u,v)	&\le&	\lambda(-R,v)e^{\int_{u}^{-R}{\frac{2}{r^2}
			\frac{\lambda|\nu|}{1-\mu}md\tilde{u}}}\\
	&\le&	e^{8\alpha^2M\int_{r(u,v)}^{-R}{\frac{dr}{r^2}}}\\
	&\le&	e^{8\alpha^2MR^{-1}}\\
	& < &	\alpha.
\end{eqnarray*}

For $m$, from $(\ref{puqu})$ we compute
\begin{eqnarray*}
m(u,v)	& = &	m(-R,v)+\int_{-R}^u{\frac12\frac{1-\mu}
			\nu\zeta^2d\tilde{u}}\\
	&\le&	M+\int_{-R}^u{\frac12\frac{1-\mu}\nu\left[
			|\zeta+\nu\phi|+|\nu\phi|\right]^2d\tilde{u}}\\
	&\le&	M+\int_{-R}^u{\frac{1-\mu}\nu\left[
			(\zeta+\nu\phi)^2+\nu^2\phi^2\right]d\tilde{u}}\\
	&\le&	M+4\alpha^2C^2\int_{-R}^u{|u|^{-3}du}	
			+4\alpha^2 C^2\int_{-R}^{r(u,v)}{\frac{dr}{r^2}}\\
	&\le&	M+2\alpha^2C^2R^{-2}+2\alpha^2 C^2R^{-1}\\
	& < &	\alpha M,
\end{eqnarray*}
where for the last inequality, we use $R>\max\left\{
6\alpha^2C^2(\alpha-1)^{-1},1\right\}$.

Finally from this, we obtain
$1-\mu=1-\frac{2m}r\ge1-\frac{4\alpha M}R > \alpha^{-1}$
since $R>4\alpha M(1-\alpha^{-1})^{-1}$.

Since 
\[
J^{-}_{-g}(\overline{\mathcal{R}})\subset \overline{J^{-}_{-g}(\mathcal{R})}=
\overline{\mathcal{R}}
\]
we have just shown that for all $(u,v)\in \overline{\mathcal{R}}$,
the inequalities $(\ref{BS1})$--$(\ref{BS6})$ hold
throughout $J^{-}_{-g}(u,v)$ with $\alpha$ replacing $2\alpha$.
Thus $(u,v)\in \mathcal{R}$, so 
\[
\mathcal{R}=\overline{\mathcal{R}}.
\]
Since ${\bf D}$ is connected, and $\mathcal{R}$ is open,
it follows that 
\[
\mathcal{R}={\bf D}.
\]

Suppose now $(u,v)\in\partial{\bf D}$. By continuity, we
have $m(u,v)\le \alpha M<\infty$, and 
\begin{eqnarray*}
r(u,v)	&\le&	R+\int_{-R}^u{\nu(\tilde{u},v)d\tilde{u}}\\
	&\le&	R+(-R-u)\alpha\\
	& < &	\infty.
\end{eqnarray*}
Thus, by the Corollary to Proposition \ref{epektasn}, 
it follows that $\partial{\bf D}=\emptyset$, and thus
\[
{\bf D}=\mathcal{R}=(-\infty,-R]\times[R,\infty).
\]

To complete the proof of the proposition,
it remains to show $(\ref{E2})$, and, under the additional assumption 
$(\ref{AA1})$, $(\ref{E6})$.
Integrating $(\ref{avafora})$ and applying
$(\ref{ud2})$, $(\ref{E1})$, $(\ref{E3})$, $(\ref{E5})$, $(\ref{E6})$, 
we obtain
\begin{eqnarray*}
|\nu\phi+\zeta|(u,v)	&\le&	|\nu\phi+\zeta|(u,R)+
				\int_R^v{\frac{|\phi|}{r^2}
				\frac{2\lambda|\nu|}{1-\mu}m
				d\tilde{v}}\\
		&\le&	C|u|^{-2}+\int_R^{r(u,v)}{\frac{|\phi|}{r^2}
				\frac{2|\nu|}{1-\mu}m
				dr}\\
		&\le&	C|u|^{-2}+2C\alpha^4M
				\int_R^{r(u,v)}{\frac{dr}{r^3}}\\
		&\le&	C|u|^{-2}+C\alpha^4Mr(u,R)^{-2}\\
		& = &	(1+\alpha^4M)C|u|^{-2}.
\end{eqnarray*}
This gives $(\ref{E2})$. On the other hand, assuming $(\ref{AA1})$
and integrating $(\ref{avafora2})$, we obtain
\begin{eqnarray*}
|\lambda\phi+\theta|(u,v)	&\le&	|\lambda\phi+\theta|(-R,v)+
				\int_{u}^{-R}{\frac{|\phi|}{r^2}
				\frac{2\lambda|\nu|}{1-\mu}m
				d\tilde{u}}\\
		&\le&	\tilde{C}v^{-2}+\int_R^{r(u,v)}{\frac{|\phi|}{r^2}
				\frac{2\lambda}{1-\mu}m
				dr}\\
		&\le&	\tilde{C}v^{-2}+2C\alpha^4M
				\int_R^{r(u,v)}{\frac{dr}{r^3}}\\
		&\le&	\tilde{C}v^{-2}+C\alpha^4Mr(-R,v)^{-2}\\
		& = &	(\tilde{C}+C\alpha^4M)v^{-2}.
\end{eqnarray*}
This completes the proof. $\Box$
 
\section{Proof of the Theorem}

In this section, we shall combine Proposition \ref{kurioprop}
with previous
results of Christodoulou to give the proof of Theorem 1.

Fix $\alpha$, $C$ and $M$, and let $R$ be as in Proposition \ref{kurioprop}.
On the ray $v=R$, $u\le0$, prescribe
\[
m(0,R)=0,
\]
\[
r=-u,
\]
and an arbitrary $\mathcal{C}^2$ function $\phi(u)$
satisfying
\begin{equation}
\label{aptovXri}
\int_{-\infty}^0{|\partial^2_u(u\phi)|du}<\epsilon_*,
\end{equation}
\begin{equation}
\label{akoma1}
|\partial_u(u\phi)|(u,R)<C|u|^{-2},
\end{equation}
for $u\ge R$,
and
\[
|R\phi(-R,R)|<C,
\]
\begin{equation}
\label{akoma2}
\int_{-\infty}^0\frac12(\partial_u\phi)^2u^2<M.
\end{equation}
If $\epsilon_*>0$ is sufficiently small, then, by Theorem 6.1 of~\cite{chr:bv}, 
$(\ref{aptovXri})$ implies that this initial value problem has
a past-causally geodesically complete asymptotically flat development:
\[
\includegraphics{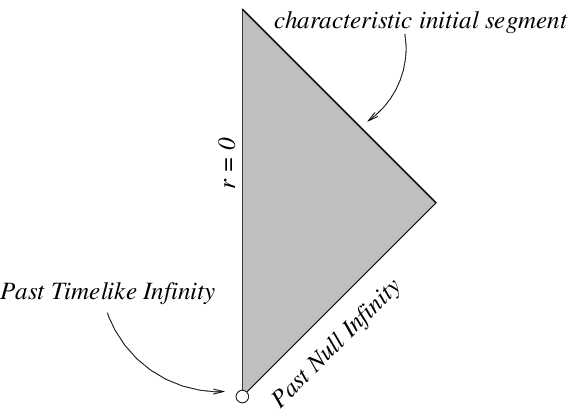}
\]
When written as a first order system on $Q$, the boundary conditions
imposed on the axis will be $r=0$, $-u+v=R$. 
Moreover, if $\phi$ is $\mathcal{C}^\infty$, then this development
is also $\mathcal{C}^\infty$.

From the above spacetime, consider the ray $u=-R$, and extend it
off the shaded region to all positive values of $v$:
\[
\includegraphics{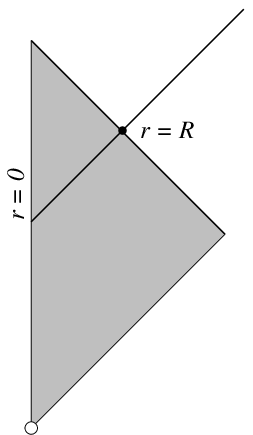}
\]
Where $v\ge R$, prescribe 
\begin{equation}
\label{kaivourgio}
r=v
\end{equation}
and
denote by $C_0=|R\phi(-R,R)|$.

Now, choose some $r_1$ satisfying $R<r_1<R+1$, and, setting $m(-R,R)=m_0$,
choose some $m_1$ satisfying $m_0<m_1<M$.
Define the function 
\[
I(V)=\min\left\{\left(\frac{C-C_0}2\right)^2\frac{1}{(V-r_1)^2},
	\frac{M-m_1}{2R^2(V-r_1)}\right\}.
\]

Fixing for the time being a $V>r_1$,
from the definition of $I$, it is clear by a partition of unity
argument that one can construct a large class of functions 
$\phi:-R\times[0,\infty)\to{\bf R}$, coinciding with
the $\phi$ induced from the solution already
obtained in $-R\times[0,R]$, and 
such that
\begin{enumerate}
\item
The function $\phi$, when viewed as a function of $r$ on $[0,\infty)$,
is regular, for instance $\mathcal{C}^\infty$. 
\item
The inequality
\begin{equation}
\label{buyuk}
(\partial_v\phi)^2(u,v)> I(V),
\end{equation}
holds for
$v\in[r_1,V]$
\item
The inequalities 
\begin{equation}
\label{sugxrovws1}
|r\phi|(-R,v)< C,
\end{equation}
\begin{equation}
\label{sugxrovws2}
m(-R,v)< M,
\end{equation}
hold
for all $v\ge R$,
where $m$ is defined by integrating $(\ref{pvqu})$.
\end{enumerate}
We will denote the class of functions satisfying properties
1, 2, and 3 above as $\mathcal{F}_V$.

Define the quantities
\[
\eta(V)=\min_{\phi\in\mathcal{F}_V}V^{-1}(m(-R,V)-m_1)
\]
and
\[
\delta(V)=Vr_1^{-1}-1.
\] 
It follows that
\begin{eqnarray*}
\eta(V)	& = &	\min_{\phi\in\mathcal{F}_V}V^{-1}\int_{r_1}^V
			{\frac12\theta^2
			\frac{1-\mu}\lambda dv}\\
	&\ge&	\min_{\phi\in\mathcal{F}_V}
			R^2V^{-1}\int_{r_1}^V{\frac12
			(\partial_v\phi)^2dv}\\
	&\ge&	\frac12R^2V^{-1}(V-r_1)I(V)\\
	&\ge&	\frac12R^2(R+1)^{-1}(V-r_1)I(V).
\end{eqnarray*}
Thus for small enough $V-r_1$ we have
\begin{equation}
\label{biryildiz}
\eta(V)\ge\frac{M-m_1}{4(R+1)}.
\end{equation}

Recall the function $E(y)$ defined in~\cite{chr:fbh} by
\[
E(y)=\frac{y}{(1+y)^2}\left[\log\left(\frac{1}{2y}\right)+5-y\right].
\]
Note that $E(y)\to0$ as $y\to0$. Choose $V-r_1$ small enough so that
\begin{equation}
\label{ikiyildiz}
E(\delta)<\frac{M-m_1}{4(R+1)}
\end{equation}
Inequalities
$(\ref{biryildiz})$ and 
$(\ref{ikiyildiz})$ together imply
\begin{equation}
\label{suv9nkn}
\eta>E(\delta).
\end{equation}
From Theorem 5.1 of~\cite{chr:fbh}, property $(\ref{suv9nkn})$ ensures
that initial data in
$\mathcal{F}_V$ has an incomplete future
development; moreover, its conformal structure is as depicted below:
\[
\includegraphics{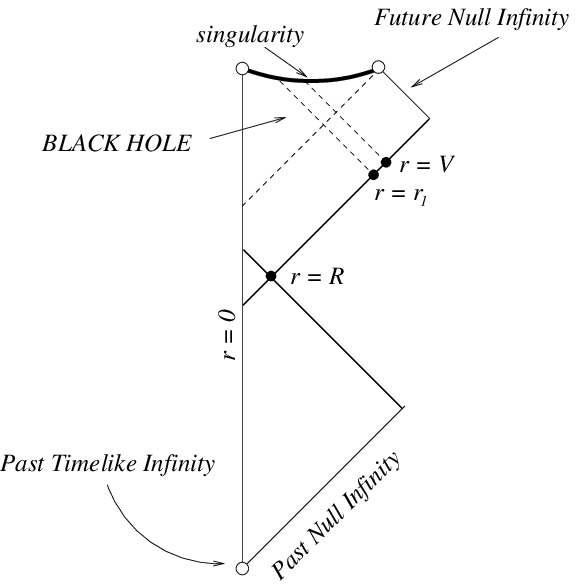}
\]
(See~\cite{chr:fbh} for details.)
On the other hand, $(\ref{akoma1})$, $(\ref{akoma2})$,
$(\ref{sugxrovws1})$, and $(\ref{sugxrovws2})$
imply that the rays $u=-R$, and $v=R$,
and the functions $r$, $\phi$, and $m$ defined on them satisfy the
conditions of Proposition \ref{kurioprop}. 
Applying this proposition one obtains a spacetime as depicted in the conformal
diagram of the Introduction. 
It is easy to see that this spacetime admits an asymptotically
flat Cauchy hypersurface. $\Box$

The strict inequalities in $(\ref{akoma1})$, $(\ref{akoma2})$,
$(\ref{sugxrovws1})$, and $(\ref{sugxrovws2})$ immediately
indicate one notion according to which our solutions are stable in the 
spherically symmetric category. Alternatively,
considering a Cauchy surface $\mathcal{S}$ intersecting $(-R,R)$,
it follows that given any solution of the kind we have constructed,
sufficiently small\footnote{For instance, in the $\mathcal{C}^\infty$ topology,
if we consider only $\mathcal{C}^\infty$ solutions.}
 perturbations supported on $\mathcal{S}$ in $r<R$ lead to a Cauchy
development with a similar conformal diagram. For on 
the backwards characteristic, $(\ref{aptovXri})$, $(\ref{akoma1})$,
and $(\ref{akoma2})$ hold on account of Cauchy stability (this takes
care of $r\le R$) and the domain of dependence theorem (for $r\ge R$),
thus guaranteeing the past completeness. On the other hand,
Cauchy stability implies that $\eta>E(\delta)$ for these purturbations,
as this refers to a compact set of the development. Thus, the formation
of a black-hole in the future is also a stable feature, in this sense.

Another point is worth mentioning. As the formation of a black hole
depends only on the solution on $-R\times[r_1,V]$, we can replace
the part of the solution for $v\ge V$
with the solution of yet another characteristic
initial value problem, where initial data are given on
$v=V$ and on past null infinity for $v\ge V$. In particular, one
can provide solutions with the conformal diagram indicated in
the Introduction, for which there is no incoming radiation near
spacelike infinity, i.e.~for which $m$ is constant along 
past null infinity in a neighborhood of spacelike infinity.
On the other hand, by the results of~\cite{chr:mt},
if past null infinity is complete and $m$ is constant throughout,
then either $m=0$, in which case the solution is Minkowski space,
or else it contains a white hole.
\newpage

\end{document}